\documentclass{article}
\usepackage{amsfonts}
\usepackage{amsmath}
\usepackage{graphics}
\usepackage{color}

\input{tcilatex}
\begin{document}

\title{\textquotedblleft Millikan oil drops\textquotedblright\ as quantum
transducers between electromagnetic and gravitational radiation}
\author{Raymond Y. Chiao \and Professor in the School of Natural Sciences
\and and in the School of Engineering \\
University of California, P. O. Box 2039\\
Merced, CA 95344 \and E-mail: rchiao@ucmerced.edu}
\date{February 25aa, 2007 Manuscript for Phys. Rev. D}
\maketitle

\begin{abstract}
Pairs of Planck-mass-scale drops of superfluid helium coated by electrons
(i.e., \textquotedblleft Millikan oil drops\textquotedblright ), when
levitated in the presence of strong magnetic fields and at low temperatures,
can be efficient quantum transducers between electromagnetic (EM) and
gravitational (GR) radiation. A Hertz-like experiment, in which EM waves are
converted at the source into GR waves, and then back-converted at the
receiver from GR waves back into EM waves, should be practical to perform.
This would open up observations of the gravity-wave analog of the Cosmic
Microwave Background from the extremely early Big Bang, and also
communications directly through the interior of the Earth.
\end{abstract}

\section{Introduction}

The problem of generating gravitational radiation in the laboratory has
historically been an exceedingly difficult one because of the extreme
smallness of Newton's constant $G$ and the extreme largeness of the speed of
light $c$. \ This paper explores situations in which the introduction of yet
another physical constant, namely Planck's constant $h$, due to its extreme
smallness, can under certain circumstances lead to the efficient generation,
as well as detection, of gravity waves. Under these circumstances, the
reciprocity principle (i.e., time-reversal symmetry) demands the existence
of nonnegligible back-actions of a quantum measuring device upon incident
gravitational radiation fields. This leads to the possibility of generating,
as well as detecting, gravity waves in a symmetric manner.

The quantum approach being suggested here is in stark contrast to the
classical, test-particle approaches being taken in contemporary, large-scale
gravity-wave experiments, which are based solely on classical physics. The
back-actions of classical measuring devices such as Weber bars and large
laser interferometers upon the incident gravitational fields that are being
measured, are completely negligible. Hence they can only detect gravity
waves from powerful astronomical sources such as supernovae \cite{MTW}, but
they certainly cannot generate these waves.

Specifically, the quantum physics of Planck-mass-scale \textquotedblleft
Millikan oil drops\textquotedblright\ consisting of electron-coated
superfluid helium drops at milli-Kelvin-scale temperatures in the presence
of Tesla-scale magnetic fields, will be explored to see if radiative
couplings mediated by such drops are large enough to be experimentally
interesting. We shall see that numerical estimates based on quantum
mechanics lead to the conclusion that efficient quantum transducers between
electromagnetic and gravitational waves consisting of pairs of such drops
should indeed be possible to implement in table-top-scale experiments. Such
experiments, some of which will be described below, have become practical to
perform because of advances in dilution refrigerator technology, and I am
planning to perform them with my colleagues at the new 10th campus of the
University of California at Merced. These experiments will explore some
interesting new physics lying at the interface of quantum mechanics and
general relativity, in which gravitational radiative effects on
macroscopically coherent, charged quantum systems, and their reciprocal
effects back on gravitational radiation fields, should become observable.

The paper is organized as follows: Sections 2 and 3 introduce the ratio of
coupling constants of gravity to electricity for two electrons in the
vacuum, which ratio also applies to the radiative powers emitted by these
electrons when they are set into motion. Section 4 introduces the Planck
mass scale in the context of the low temperature physics of macroscopically
coherent, charged quantum matter. Maxwell-like equations for weak
gravitational radiation fields that result from linearizing Einstein's
equations are introduced in Section 5, and the boundary conditions and the
solutions of these equations for the reflections of gravity waves from
superconductors and quantum Hall fluids are discussed. In Sections 7 and 8,
\textquotedblleft Millikan oil drops\textquotedblright\ are described in
more detail, and their role as quantum transducers in a Hertz-like
experiment to generate and detect gravitational radiation is discussed.\
Section 9 examines the quantum conditions for these drops to exhibit a M\"{o}%
ssbauer-like response to radiation fields. In Section 10, time-reversal
symmetry is applied to the differential scattering cross-section, and an
estimation of the total cross-section based on the Maxwell-like equations
introduced in Section 5 is made for a pair of Mie-resonant drops. A common
misconception that the drops are too heavy to move appreciably, which
ignores the role of the equivalence principle, is corrected in Section 11.
The final three sections present numerical considerations to show the
feasibility of the Hertz-like experiment, and discuss its possible
applications.

\section{Forces of gravity and electricity between two electrons}

Let us first consider, using only classical, Newtonian concepts (which are
valid in the correspondence-principle limit and at large distances
asymptotically, as seen by a distant observer), the forces experienced by
two electrons separated by a distance $r$ in the vacuum. Both the
gravitational and the electrical force obey long-range, inverse-square laws.
Newton's law of gravitation states that%
\begin{equation}
\left\vert F_{G}\right\vert =\frac{Gm_{e}^{2}}{r^{2}}
\label{Newton's-inverse-square-law}
\end{equation}%
where $G$ is Newton's constant and $m_{e}$ is the mass of the electron.
Coulomb's law states that%
\begin{equation}
\left\vert F_{e}\right\vert =\frac{e^{2}}{r^{2}}\text{ }
\label{Coulomb's-law}
\end{equation}%
where $e$ is the charge of the electron. The electrical force is repulsive,
and the gravitational one attactive.

Taking the ratio of these two forces, one obtains the dimensionless ratio of
coupling constants%
\begin{equation}
\frac{\left\vert F_{G}\right\vert }{\left\vert F_{e}\right\vert }=\frac{%
Gm_{e}^{2}}{e^{2}}\approx 2.4\times 10^{-43}\text{ .}  \label{Gm^2/e^2}
\end{equation}%
The gravitational force is extremely small compared to the electrical force,
and is therefore usually omitted in all treatments of quantum physics.

\section{Gravitational and electromagnetic radiation powers emitted by two
electrons}

The above ratio of the coupling constants $Gm_{e}^{2}/e^{2}$ is also the
ratio of the powers of gravitational (GR) to electromagnetic (EM) radiation
emitted by two electrons separated by a distance $r$ in the vacuum, when
they undergo an acceleration $a$ and are moving with a speed $v$ relative to
each other, as seen by a distant observer.

From the equivalence principle, it follows that dipolar gravitational
radiation does not exist. \ Rather, the lowest order radiation permitted by
this principle is quadrupolar, and not dipolar, in nature. General
relativity predicts that the power $P_{GR}^{\text{(quad)}}$\ radiated by a
time-varying mass quadrupole tensor $D_{ij}$ of a periodic system is given
by \cite{Landau}\cite{Weinberg}%
\begin{equation}
P_{GR}^{\text{(quad)}}=\frac{G}{45c^{5}}\left\langle \dddot{D}%
_{ij}^{2}\right\rangle =\omega ^{6}\frac{G}{45c^{5}}\left\langle
D_{ij}^{2}\right\rangle  \label{triple-dot}
\end{equation}%
where the triple dots over $\dddot{D}_{ij}$ denote the third derivative with
respect to time of the mass quadrupole-moment tensor $D_{ij}$ of the system
(the Einstein summation convention over the spatial indices $(i,j)$ for the
term $\dddot{D}_{ij}^{2}$ is being used here), $\omega $ is the angular
frequency of the periodic motion of the system, and the angular brackets
denote time averaging over one period of the motion.

Applying this formula to the periodic orbital motion of two point masses
with equal mass $m$ moving with a relative instantaneous acceleration whose
magnitude is given by $\left\vert a\right\vert =\omega ^{2}\left\vert
D\right\vert $, where $\left\vert D\right\vert $ is the magnitude of the
relative displacement of these objects, and where the relative instantaneous
speed of the two masses is given by $\left\vert v\right\vert =\omega
\left\vert D\right\vert $ (where $v<<c$), all these quantities being
measured by a distant observer, one finds that Equation (\ref{triple-dot})
can be rewritten as follows:%
\begin{equation}
P_{GR}^{\text{(quad)}}=\kappa \frac{2}{3}\frac{Gm^{2}}{c^{3}}a^{2}\text{
where }\kappa =\frac{2}{15}\frac{v^{2}}{c^{2}}.  \label{modified-Larmor}
\end{equation}%
The frequency dependence of the radiated power predicted by Equation (\ref%
{modified-Larmor}) scales as $v^{2}a^{2}\sim \omega ^{6}$, in agreement with
triple dot term $\dddot{D}_{ij}^{2}$ in Equation (\ref{triple-dot}). It
should be stressed that the values of the quantities $a$ and $v$ are those
being measured by an observer at infinity. The validity of Equations (\ref%
{triple-dot}) and (\ref{modified-Larmor}) has been verified by observations
of the orbital decay of the binary pulsar PSR 1913+16 \cite{Taylor1994}.

Now consider the radiation emitted by two electrons undergoing an
acceleration $a$ relative to each other with a relative speed $v$, as
observed by an observer at infinity. For example, these two electrons could
be attached to the two ends of a massless, rigid rod rotating around the
center of mass of the system like a dumbbell. The power in gravitational
radiation that they will emit is given by%
\begin{equation}
P_{GR}^{\text{(quad)}}=\kappa \frac{2}{3}\frac{Gm_{e}^{2}}{c^{3}}a^{2}\text{,%
}  \label{GR-Larmor}
\end{equation}%
where the factor $\kappa $ is given above in Equation (\ref{modified-Larmor}%
). Due to their bilateral symmetry, these two identical electrons will also
radiate quadrupolar, but not dipolar, electromagnetic radiation with a power
given by%
\begin{equation}
P_{EM}^{\text{(quad)}}=\kappa \frac{2}{3}\frac{e^{2}}{c^{3}}a^{2}\text{ ,}
\label{EM-Larmor-quad}
\end{equation}%
with the same factor of $\kappa $. The reason that this is true is that any
given electron carries with it mass as well as charge as it moves, since its
charge and mass must co-move rigidly together. Therefore two electrons
undergoing an acceleration $a$ relative to each other with a relative speed $%
v$ will emit simultaneously both electromagnetic and gravitational
radiation, and the quadrupolar electromagnetic radiation which it emits will
be completely homologous to the quadrupolar gravitational radiation which it
also emits.

It follows that the ratio of gravitational to electromagnetic radiation
powers emitted by the two-electron system is given by the same ratio of
coupling constants as that for the force of gravity relative to the force of
electricity, viz.,%
\begin{equation}
\frac{P_{GR}^{\text{(quad)}}}{P_{EM}^{\text{(quad)}}}=\frac{Gm_{e}^{2}}{e^{2}%
}\approx 2.4\times 10^{-43}\text{ .}  \label{Ratio-of-powers}
\end{equation}%
Thus it would seem at first sight to be hopeless to try and use any
two-electron system as the means for coupling between electromagnetic and
gravitational radiation.

Nevertheless it must be emphasized here that although this dimensionless
ratio of coupling constants is extremely small, the gravitational radiation
emitted from the two electron system must \emph{in principle} exist, or else
there must be something fundamentally wrong with the experimentally
well-tested inverse-square laws given by Equations (\ref%
{Newton's-inverse-square-law}) and (\ref{Coulomb's-law}).

\section{The Planck mass scale}

\begin{figure}[tbp]
\centerline{\includegraphics{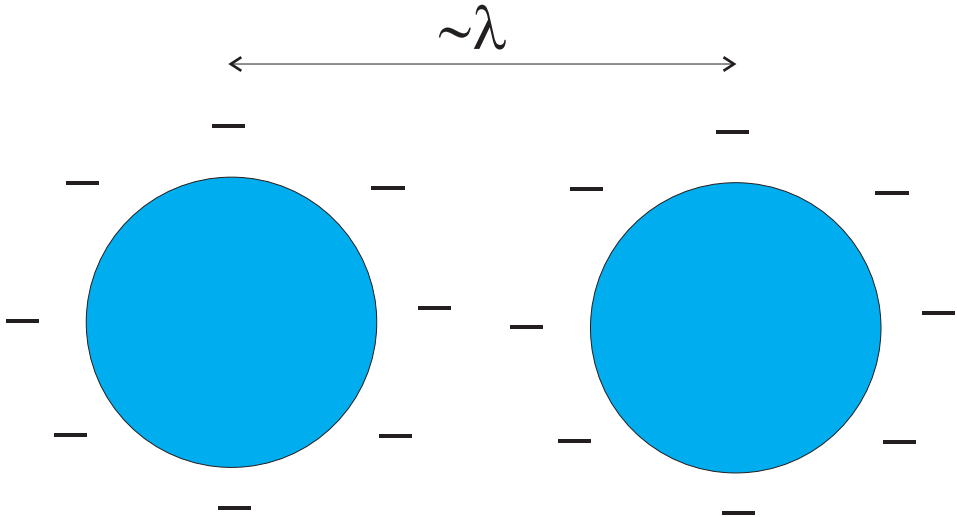}}
\caption{Planck-mass-scale superfluid helium drops coated with electrons on
their outside surfaces and separated by around a microwave wavelength $%
\protect\lambda $, which are levitated in the presence of a strong magnetic
field (not to scale).}
\label{2-oil-drops-in-trap-4b}
\end{figure}

However, the ratio of the forces of gravity and electricity of two
\textquotedblleft Millikan oil drops\textquotedblright\ (to be described in
more detail below; see Figure \ref{2-oil-drops-in-trap-4b}) needs not be so
hopelessly small \cite{Lamb-medal}.

Suppose that each \textquotedblleft Millikan oil drop\textquotedblright\ has
a single electron attached firmly to it,\ and contains a Planck-mass amount
of superfluid helium, viz.,%
\begin{equation}
m_{\text{Planck}}=\sqrt{\frac{\hbar c}{G}}\approx 22\text{ micrograms}
\label{Planck-mass}
\end{equation}%
where $\hbar $ is Planck's constant/2$\pi $, $c$ is the speed of light, and $%
G$ is Newton's constant. Planck's mass sets the characteristic scale at
which quantum mechanics ($\hbar $) impacts relativistic gravity ($c$, $G$).
(Planck obtained this mass by means of dimensional analysis.) Note that the
extreme smallness of $\hbar $ compensates for the extreme largeness of $c$
and for the extreme smallness of $G$, so that this mass scale is \textit{%
mesoscopic}, and not astronomical, in size. This suggests that it may be
possible to perform some novel \textit{nonastronomical}, table-top-scale
experiments at the interface of quantum mechanics and general relativity,
which are accessible in the laboratory. Such experiments will be considered
here.

The forces of gravity and electricity between the two \textquotedblleft
Millikan oil drops\textquotedblright\ are exerted upon the centers of mass
and the centers of charge of the drops, respectively. Both of these centers
coincide with the geometrical centers of the spherical drops, assuming that
the charge of the electrons on the drops is uniformly distributed around the
outside surface of the drops in a spherically symmetric manner (like in an $%
S $ state). Therefore the ratio of the forces of gravity and electricity
between the two \textquotedblleft Millikan oil drops\textquotedblright\ now
becomes%
\begin{equation}
\frac{\left\vert F_{G}\right\vert }{\left\vert F_{e}\right\vert }=\frac{Gm_{%
\text{Planck}}^{2}}{e^{2}}=\frac{G\left( \hbar c/G\right) }{e^{2}}=\frac{%
\hbar c}{e^{2}}\approx 137\text{ .}  \label{137}
\end{equation}%
Now the force of gravity is approximately 137 times stronger than the force
of electricity, so that instead of a mutual repulsion between these two
charged, massive objects, there is now a mutual attraction between them. The
sign change from mutual repulsion to mutual attraction between these two
\textquotedblleft Millikan oil drops\textquotedblright\ occurs at a critical
mass $m_{\text{crit}}$ given by%
\begin{equation}
m_{\text{crit}}=\sqrt{\frac{e^{2}}{\hbar c}}m_{\text{Planck}}\approx 1.9%
\text{ micrograms}  \label{m_[crit]}
\end{equation}%
whereupon $\left\vert F_{G}\right\vert $ $=\left\vert F_{e}\right\vert $,
and the forces of gravity and electricity balance each other in equilibrium.
The radius of a drop with this critical mass of superfluid helium, which has
a density of $\rho =0.145$ g/cm$^{3}$, is%
\begin{equation*}
R=\left( \frac{3m_{\text{crit}}}{4\pi \rho }\right) ^{1/3}=146\text{
micrometers.}
\end{equation*}%
This is a strong hint that mesoscopic-scale quantum effects can lead to
non-negligible couplings between gravity and electromagnetism that can be
observed in the laboratory.

The critical mass $m_{\text{crit}}$ is also the mass at which there occurs a
comparable amount of generation of electromagnetic and gravitational
radiation power upon scattering of radiation from the pair of
\textquotedblleft Millikan oil drops,\textquotedblright\ each with a mass $%
m_{\text{crit}}$ and with a single electron attached to it. The ratio of
quadrupolar gravitational to the quadrupolar electromagnetic radiation power
is given by%
\begin{equation}
\frac{P_{GR}^{\text{(quad)}}}{P_{EM}^{\text{(quad)}}}=\frac{Gm_{\text{crit}%
}^{2}}{e^{2}}=1\text{ ,}  \label{Larmor-power-ratio}
\end{equation}%
where the factors of $\kappa $ in Equations (\ref{GR-Larmor}) and (\ref%
{EM-Larmor-quad}) cancel out, if the center of mass of the drop co-moves
rigidly together with its center of charge. This implies that the scattered
power from these two charged objects in the gravitational wave channel will
become equal to that in the electromagnetic wave channel. Note that a pair
of larger drops, whose masses have been increased beyond the critical mass,
will still satisfy Equation (\ref{Larmor-power-ratio}), provided that the
number of electrons on these drops is also increased proportionately so that
the charge-to-mass ratio of these drops remains fixed, and provided that the
system is placed in a strong magnetic field and cooled to low temperatures
so that it remains in the ground state.

\section{Maxwell-like equations that result from linearizing Einstein's
equations}

In order to understand the calculation of the scattering cross section of
the \textquotedblleft Millikan oil drops\textquotedblright\ to be given
below, let us start from a very useful Maxwell-like representation of the
linearized Einstein's equations of standard general relativity that
describes weak gravitational fields coupled to matter in the asymptotically
flat coordinate system of a distant inertial observer \cite{Wald}: 
\begin{equation}
\mathbf{\nabla \cdot E}_{G}=-\frac{\rho _{G}}{\varepsilon _{G}}
\label{Maxwell-like-eq-1}
\end{equation}%
\begin{equation}
\mathbf{\nabla \times E}_{G}=-\frac{\partial \mathbf{B}_{G}}{\partial t}
\label{Maxwell-like-eq-2}
\end{equation}%
\begin{equation}
\mathbf{\nabla \cdot B}_{G}=0  \label{Maxwell-like-eq-3}
\end{equation}%
\begin{equation}
\mathbf{\nabla \times B}_{G}=\mu _{G}\left( -\mathbf{j}_{G}+\varepsilon _{G}%
\frac{\partial \mathbf{E}_{G}}{\partial t}\right)  \label{Maxwell-like-eq-4}
\end{equation}%
where the gravitational analog of the magnetic permeability of free space is
given by%
\begin{equation}
\mu _{G}=\frac{4\pi G}{c^{2}}=9.31\times 10^{-27}\text{ SI units}
\label{mu_G}
\end{equation}%
and where the gravitational analog of the electric permittivity of free
space is given by%
\begin{equation}
\varepsilon _{G}=\frac{1}{4\pi G}=1.19\times 10^{9}\text{ SI units.}
\label{epsilon_G}
\end{equation}%
Taking the curl of the gravitational analog of Faraday's law, Equation (\ref%
{Maxwell-like-eq-2}), and substituting into its right side the gravitational
analog of Ampere's law, Equation (\ref{Maxwell-like-eq-4}), one obtains a
wave equation, which implies that the speed of gravitational radiation is
given by%
\begin{equation}
c=\frac{1}{\sqrt{\varepsilon _{G}\mu _{G}}}=3.00\times 10^{8}\text{ m/s,}
\label{speed-of-light}
\end{equation}%
which exactly equals the vacuum speed of light. In these Maxwell-like
equations, the field $\mathbf{E}_{G}$, which is the \emph{gravito-electric}
field, is to be identified with the local acceleration $\mathbf{g}$ of a
test particle produced by the mass density $\rho _{G}$, and the field $%
\mathbf{B}_{G}$, which is the \emph{gravito-magnetic} field produced by the
mass current density $\mathbf{j}_{G}$ and by the gravitational analog of the
Maxwell displacement current density $\varepsilon _{G}\partial \mathbf{E}%
_{G}/\partial t$, is to be identified with the Lense-Thirring field of
general relativity.

In addition to the speed $c$ of gravity waves, there is another important
physical property that these waves possess, which can be formed from the
gravito-magnetic permeability of free space $\mu _{G}$ and from the
gravito-electric permittivity $\varepsilon _{G}$ of free space, namely, the
characteristic impedance of free space $Z_{G}$, which is given by \cite%
{Kiefer-Weber}\cite{Chiao2004}%
\begin{equation}
Z_{G}=\sqrt{\frac{\mu _{G}}{\varepsilon _{G}}}=2.79\times 10^{-18}\text{ SI
units.}  \label{Z_G}
\end{equation}%
As in electromagnetism, this characteristic impedance of free space plays an
important role in all radiation problems, such as in a comparison of the
radiation resistance of gravity-wave antennas to the value of this impedance
in order to estimate the coupling efficiency of these antennas to free
space. The numerical value of this impedance is extremely small, but the
impedance of all material objects must be \textquotedblleft impedance
matched\textquotedblright\ to this extremely small quantity before
significant power can be transferred efficiently from gravity waves to these
objects, or vice versa.

However, all classical material objects, such as Weber bars, have such a
high dissipation and such a high radiation resistance that they are
extremely poorly impedance-matched to free space. They can therefore neither
absorb gravity wave energy, nor emit it efficiently \cite{Weinberg}\cite%
{Chiao2004}. Hence it is a common belief that all materials, whether
classical or quantum, are essentially completely transparent to
gravitational radiation. Quantum materials, such as superconductors,
however, can be exceptions to this general rule, since they can have a
strictly zero dissipation which arises from an energy gap (the BCS gap) in
the material, as evidenced by the persistent currents in annular rings of
superconductors that have been projected to last longer than the age of the
Universe \cite{Tinkham}.

\section{Specular reflection of gravity waves from superconductors and other
quantum fluids}

One important consequence of the zero-resistance property of superconductors
is that a mirror-like reflection should occur at a planar
vacuum-superconductor interface. This reflection is similar to that which
occurs when an incident electromagnetic wave propagates down a transmission
line with a characteristic impedance $Z$, which is then terminated by means
of a resistor $R$ whose value is close to zero. The reflection coefficient ${%
\mathcal{R}}$ of the wave from such a termination is given by%
\begin{equation}
{\mathcal{R}}=\left\vert \frac{Z-R}{Z+R}\right\vert ^{2}\rightarrow 100\%%
\text{ when }R\rightarrow 0\text{ ,}
\label{Reflection-from-transmission-line}
\end{equation}%
which approaches arbitrarily close to 100\% when the resistance vanishes.

From the Maxwell-like Equations (\ref{Maxwell-like-eq-1}) - (\ref%
{Maxwell-like-eq-4}), and the boundary conditions that follow from them, it
follows that there should exist an analogous reflection of a gravitational
plane wave from a planar vacuum-superconductor interface, whose Fresnel-like
reflection coefficient ${\mathcal{R}}_{G}$ is given by%
\begin{equation}
{\mathcal{R}}_{G}=\left\vert \frac{Z_{G}-R_{G}}{Z_{G}+R_{G}}\right\vert
^{2}\rightarrow 100\%\text{ when }R_{G}\rightarrow 0\text{ .}
\label{Reflection-from-vacuum-superconductor-interface}
\end{equation}%
This counter-intuitive result is due to the fact that the superconductor
possesses a strictly zero dissipation, and therefore an equivalent
mass-current resistance $R_{G}$ that should also be strictly zero, as
compared to the characteristic impedance of free space $Z_{G}$ $=2.79\times
10^{-18}$ SI units given by Equation (\ref{Z_G}).

Although the gravitational impedance of free space $Z_{G\text{ }}$ is an
extremely small quantity, it is still a finite quantity. However, the
resistance of a superconductor is strictly zero. Arbitrarily long-lasting
persistent mass currents in annular rings of superconductors that accompany
long-lasting persistent electrical currents are the evidence for this
experimental fact, which can be understood theoretically on a microscopic
basis using the BCS theory of superconductivity \cite{Tinkham}.

Hence it follows that nearly perfect reflection of gravity waves should
occur from a superconductor at temperatures well below its transition
temperature. Therefore nearly perfect mirrors for gravitational radiation in
principle should exist. Curved superconducting mirrors can focus this
radiation, and Newtonian telescopes for gravity waves can therefore in
principle be constructed. In the case of scattering of gravity waves from
superconducting spheres, the above specular-reflection condition implies
hard-wall boundary conditions at the surfaces of these spheres, so that the
scattering cross section of these waves from a pair of large spheres should
be geometric, i.e., hard-sphere, in size.

Not only superconductors, but also quantum Hall fluids moving frictionlessly
on a plane, could possibly exhibit nearly perfect reflectivity to normally
incident gravitational plane waves. This follows from the fact that these
fluids in their quantum Hall plateaus are also perfectly dissipationless,
just like superconductors \cite{Prange}. \ For example, in the integer
effect, the transverse quantum Hall resistance is quantized in integer
multiples of $h/e^{2}$, where $h$ is Planck's constant and $e$ is the
electron charge, but the longitudinal quantum Hall resistance,\ which is
responsible for dissipation, is strictly zero. This arises from the energy
gap of the quantum Hall system, which is proportional to the cyclotron
frequency of the electrons in the Tesla-scale magnetic fields used in
quantum Hall experiments.

However, one cannot tell whether these statements about specular reflection
of gravitational radiation from superconductors or quantum Hall fluids are
true or not experimentally, without the existence of a source and a detector
for such radiation. The quantum transducers based on \textquotedblleft
Millikan oil drops\textquotedblright\ to be discussed in more detail below
may provide the needed source and detector.

The counter-intuitive prediction of specular reflection of gravity waves
from superconductors may possibly also follow from the recent potentially
very important discovery \cite{Tajmar} (which of course needs independent
confirmation)\ that in an angularly accelerating superconductor, such as a
niobium ring rotating with a steadily increasing angular velocity, there
seems to be a large enhancement of the gravito-magnetic field, perhaps
arising from a macroscopically constructive quantum interference effect\ due
to the macroscopically coherent nature of the quantum mechanical phase of
the electrons in niobium, which in turn arises from the condensate of many
Cooper pairs of electrons in this superconductor. As a result of the angular
acceleration of the niobium ring, there seems to arise a steadily increasing
gravitational analog of the London moment in the form of a very large $%
\mathbf{B}_{G}$ field inside the ring, which is increasing linearly in time.
The gravitational analog of Faraday's law, Equation (\ref{Maxwell-like-eq-2}%
), then implies the generation of loops of the gravito-electric field $%
\mathbf{E}_{G}$ inside the hole of the ring, which can be detected by
sensitive accelerometers. The gravito-magnetic field $\mathbf{B}_{G}$ is
thus inferred to be many orders of magnitude greater than what one would
expect classically due to the mass current associated with the rigid
rotation of the ionic lattice of the superconducting ring. These
observations have recently been confirmed by replacing the electromechanical
accelerometers with a laser gyro \cite{Tajmar2}.

A tentative theoretical interpretation of these recent experiments is that
the coupling constant $\mu _{G}$ which couples the mass currents of the
superconductor to the gravito-magnetic field $\mathbf{B}_{G}$ is somehow
greatly enhanced due to the presence of the macroscopically coherent quantum
matter in niobium. This enhancement can be understood phenomenologically in
terms of a gravito-magnetic enhancement factor $\kappa _{G}^{\text{(magn)}}$%
, which enhances the gravito-magnetic coupling constant \emph{inside the
medium} as follows:%
\begin{equation}
\mu _{G}^{\prime }=\kappa _{G}^{\text{(magn)}}\mu _{G}  \label{mu-G-prime}
\end{equation}%
where $\kappa _{G}^{\text{(magn)}}$ is a positive number much larger than
unity. This gravito-magnetic enhancement factor $\kappa _{G}^{\text{(magn)}}$
is the gravitational analog of the magnetic permeability constant $\kappa
_{m}$ of ferromagnetic materials in the standard theory of electromagnetism.

The basic assumption of this phenomenological theory is that of a \emph{%
linear response} of the material medium to weak applied gravito-magnetic
fields \cite{Kramers-Kronig}; that is to say, whatever the fundamental
explanation is of the large observed positive values of $\kappa _{G}^{\text{%
(magn)}}$, the medium produces an enhanced gravito-magnetic field $\mathbf{B}%
_{G}$\ that is \emph{directly proportional} to the mass current density $%
\mathbf{j}_{G}$ arising from the motion of the ionic lattice of the
superconductor. For weak fields, this is a reasonable assumption. However,
it should be noted that this phenomenological explanation based on Equation (%
\ref{mu-G-prime}) is different from the theoretical explanation based on
Proca-like equations for gravitational fields with a finite graviton rest
mass, which was proposed by the discoverers of the effect in Ref. \cite%
{Tajmar}.

Nevertheless, it is natural to consider introducing the phenomenological
Equation (\ref{mu-G-prime}) to explain the observations, since a large
enhancement factor $\kappa _{G}^{\text{(magn)}}$\ due to the material medium
is similar to its analog in ordinary magnetism, which explains, for example,
the large ferromagnetic enhancement of the inductance of a solenoid by a
magnetically soft, permeable iron core with permeability $\kappa _{m}>>1$
that arises from the alignment of electron spins inside the iron. This
spin-alignment effect leads to the large observed values of the magnetic
susceptibility of iron. Just as in the case of the iron core inserted inside
a solenoid, where the large enhancement of the solenoid's inductance
disappears above the Curie temperature of iron, it was observed in these
recent experiments that the large gravito-magnetic enhancement effect
disappears above the superconducting transition temperature of niobium.

If the tentative phenomenological interpretation given by Equation (\ref%
{mu-G-prime}) of these experiments turns out to be correct, one important
consequence of the large resulting values of $\kappa _{G}^{\text{(magn)}}$
is that again a mirror-like reflection should occur at a planar
vacuum-superconductor interface, where the refractive index of the
superconductor has an abrupt jump from unity to a value given by%
\begin{equation}
n_{G}=\left( \kappa _{G}^{\text{(magn)}}\right) ^{1/2}\text{ .}
\label{refractive-index}
\end{equation}

However, it should be immediately emphasized here that only positive masses
are observed to exist in nature, and not negative ones. Hence gravitational
analogs of electric dipole moments do not exist. It follows that the
gravitational analog $\kappa _{G}^{\text{(elec)}}$ of the usual dielectric
constant $\kappa _{e}$ for all kinds of matter, whether classical or
quantum, in the Earth's gravito-electric field $\mathbf{E}_{G}=\mathbf{g}$,
cannot differ from its vacuum value of unity, i.e.,%
\begin{equation}
\kappa _{G}^{\text{(elec)}}\equiv \varepsilon _{G}^{\prime }/\varepsilon
_{G}=1\text{ , }  \label{gravito-dielectric-constant=1}
\end{equation}%
exactly. Hence one cannot screen out, even partially, the gravito-electric
DC gravitational fields like the Earth's gravitational field using
superconducting Faraday cages. In particular, the local value of the
acceleration $\mathbf{g}$ due to Earth's gravity is not at all affected by
the presence of nearby matter with large $\kappa _{G}^{\text{(magn)}}$, a
fact which follows from the Maxwell-like Equations (\ref{Maxwell-like-eq-1})
- (\ref{Maxwell-like-eq-4}).

By contrast, the gravitational analog of Ampere's law combined with the
analog of the Lorentz force law%
\begin{equation}
\mathbf{F}_{G}=m\left( \mathbf{E}_{G}+4\mathbf{v\times B}_{G}\right) \text{ ,%
}  \label{Loretntz-Force-Law}
\end{equation}%
where $\mathbf{F}_{G}$ is the force on a test particle with mass $m$ and
velocity $\mathbf{v}$ (all quantities as seen by the distant inertial
observer), leads to the fact that a \textit{repulsive} component of force
exists between two parallel mass currents travelling in the same direction,
which is the opposite to the case in electricity, where two parallel
electrical currents travelling in the same direction \textit{attract} each
other \cite{Wald}. A repulsive \textit{gravito-magnetic} gravitational force
necessarily follows from the negative sign in front of the mass current
density $\mathbf{j}_{G}$ in Equation (\ref{Maxwell-like-eq-4}), which is
necessitated by the conservation of mass, since upon taking the divergence
of Equation (\ref{Maxwell-like-eq-4}), and combining it with Equation (\ref%
{Maxwell-like-eq-1}) (whose negative sign in front of the mass density $\rho
_{G}$\ is fixed by Newton's law of gravitation, where all masses \emph{%
attract} each other), one must obtain the continuity equation for mass, i.e.,%
\begin{equation}
\mathbf{\nabla \cdot j}_{G}+\frac{\partial \rho _{G}}{\partial t}=0\text{ ,}
\label{continuity}
\end{equation}%
where $\mathbf{j}_{G}$ is the mass current density, and $\rho _{G}$\ is the
mass density. Moreover, the negative sign in front of the mass current
density $\mathbf{j}_{G}$ in the gravitational analog of Ampere's law,
Equation (\ref{Maxwell-like-eq-4}), implies an \textit{anti-Meissner}
effect, in which the lines of the $\mathbf{B}_{G}$\ field, instead of being
expelled from the superconductor, as in the usual Meissner effect, are
pulled tightly into the interior of the body of the superconductor.

However, it should again be stressed that what is being proposed here in the
above phenomenological scenario does not at all imply an \textquotedblleft
anti-gravity\textquotedblright\ effect, in which the Earth's gravitational
field is somehow partially screened out by the so-called \textquotedblleft
Podkletnov effect\textquotedblright , where it was claimed that rotating
superconductors reduce by a few percent the gravito-electric field $\mathbf{E%
}_{G}=\mathbf{g}$, i.e., the local acceleration of all objects due to
Earth's gravity, in their vicinity. Experiments attempting to reproduce this
effect have failed to do so\ \cite{Tajmar}. The non-existence of the
\textquotedblleft Podkletnov effect\textquotedblright\ would be consistent
with the above phenomenological theory, since \textit{longitudinal}
gravito-electric fields cannot be screened under any circumstances; however, 
\textit{transverse} radiative gravitational fields can be reflected by
macroscopically coherent quantum matter.

Very large values of $\kappa _{G}^{\text{(magn)}}$ for superconductors would
imply that the index of refraction for gravitational plane waves in these
media would be considerably larger than unity, i.e.,%
\begin{equation}
n_{G}=\left( \kappa _{G}^{\text{(magn)}}\right) ^{1/2}>>1\text{ .}
\label{index>>1}
\end{equation}%
The Fresnel reflection coefficient ${\mathcal{R}}_{G}$ of gravity waves
normally incident upon the vacuum-superconductor interface would therefore
become%
\begin{equation}
{\mathcal{R}}_{G}=\left\vert \frac{n_{G}-1}{n_{G}+1}\right\vert
^{2}\rightarrow 100\%\text{,}  \label{Fresnel-reflection}
\end{equation}%
which would again approach unity. Again, this would imply specular
reflection of these waves from superconducting surfaces. It should be noted
that large values of the gravito-magnetic permeability-enhancement factor $%
\kappa _{G}^{\text{(magn)}}$, of the index of refraction $n_{G}$, and of the
reflectivity ${\mathcal{R}}_{G}\,$, are not forbidden by the principle of
equivalence.

\section{\textquotedblleft Millikan oil drops\textquotedblright\ described
in more detail}

Let the oil of the classic Millikan oil drops be replaced with superfluid
helium ($^{4}$He) with a gravitational mass of around the
Planck-mass--scale, and let these drops be levitated in the presence of
strong, Tesla-scale magnetic fields.

The helium atom is diamagnetic, and liquid helium drops have successfully
been magnetically levitated in an anti-Helmholtz magnetic trapping
configuration \cite{Weilert1996}. Due to its surface tension, the surface of
a freely suspended, isolated, ultracold superfluid drop is atomically
perfect, in the sense that there are no defects (such as dislocations in an
imperfect crystal) which can trap and thereby localize the electron. When an
electron approaches a drop, the formation of an image charge inside the
dielectric sphere of the drop causes the electron to be attracted by the
Coulomb force to its own image. As a result, it is experimentally observed
that the electron is bound\ to the outside surface of the drop in a
hydrogenic ground state. The binding energy of the electron to the surface
of liquid helium has been measured using millimeter-wave spectroscopy to be
8 Kelvin \cite{Grimes2}, which is quite large compared to the
milli-Kelvin-scale temperatures for the proposed experiments. Hence the
electron is tightly bound to the outside surface of the drop so that the
radial component of its motion is frozen, but it is free to move
tangentially and thus to become delocalized over the surface.

Such a \textquotedblleft Millikan oil drop\textquotedblright\ is a
macroscopically phase-coherent quantum object. In its ground state, which
possesses a single, coherent quantum mechanical phase throughout the
interior of the superfluid \cite{Footnote-A}, the drop possesses a zero
circulation quantum number (i.e., contains no quantum vortices), with one
unit (or an integer multiple) of the charge quantum number. As a result of
the drop being at ultra-low temperatures, all degrees of freedom other than
the center-of-mass degrees of freedom are frozen out, so that there results
a zero-phonon M\"{o}ssbauer-like effect, in which the entire mass of the
drop moves rigidly as a single unit in response to radiation fields (see
below). Therefore, the center of mass of the drop will co-move with the
center of charge. Also, since it remains adiabatically in the ground state
during perturbations due to these weak radiation fields, the
\textquotedblleft Millikan oil drop\textquotedblright\ possesses properties
of \textquotedblleft quantum rigidity\textquotedblright\ and
\textquotedblleft quantum dissipationlessness\textquotedblright\ that are
the two most important quantum properties for achieving a high coupling
efficiency for gravity-wave antennas \cite{Chiao2004}.

Note that two spatially separated \textquotedblleft Millikan oil
drops\textquotedblright\ with the same mass and charge\ have the correct
bilateral symmetry in order to couple to quadrupolar gravitational
radiation, as well as to quadrupolar electromagnetic radiation. The coupling
of the drops to dipolar electromagnetic radiation, however, vanishes due to
symmetry. When they are separated by a distance on the order of a
wavelength, they should become an efficient quadrupolar antenna capable of
generating, as well as detecting, gravitational radiation.

\section{A pair of ``Millikan oil drops'' as a transducer}

Now imagine placing a pair of levitated \textquotedblleft Millikan oil
drops\textquotedblright\ separated by approximately a microwave wavelength
inside a black box, which represents a quantum transducer that can convert
gravitational (GR) waves into electromagnetic (EM) waves. See Figure \ref%
{GR-to-EM-wave-transducer-4b}. This kind of transducer action is similar to
that of the tidal force of a gravity wave passing over a pair of charged,
freely falling objects orbiting the Earth, which can in principle convert a
GR wave into an EM\ wave \cite{Lamb-medal}. Such transducers are linear,
reciprocal devices.

By time-reversal symmetry, the reciprocal process, in which another pair of
\textquotedblleft Millikan oil drops\textquotedblright\ converts an EM wave
back into a GR wave, must occur with the same efficiency as the forward
process, in which a GR wave is converted into an EM wave by the first pair
of \textquotedblleft Millikan oil drops.\textquotedblright\ The
time-reversed process is important because it allows the \emph{generation}
of gravitational radiation, and can therefore become a practical source of
such radiation. The radiation reaction or back-action by the EM fields upon
the GR fields via these coherent quantum mechanical drops leads necessarily
to a non-negligible reciprocal process of the generation of these fields.
These actions must be mutual ones between these two kinds of radiation
fields. 
\begin{figure}[tbp]
\centerline{\includegraphics{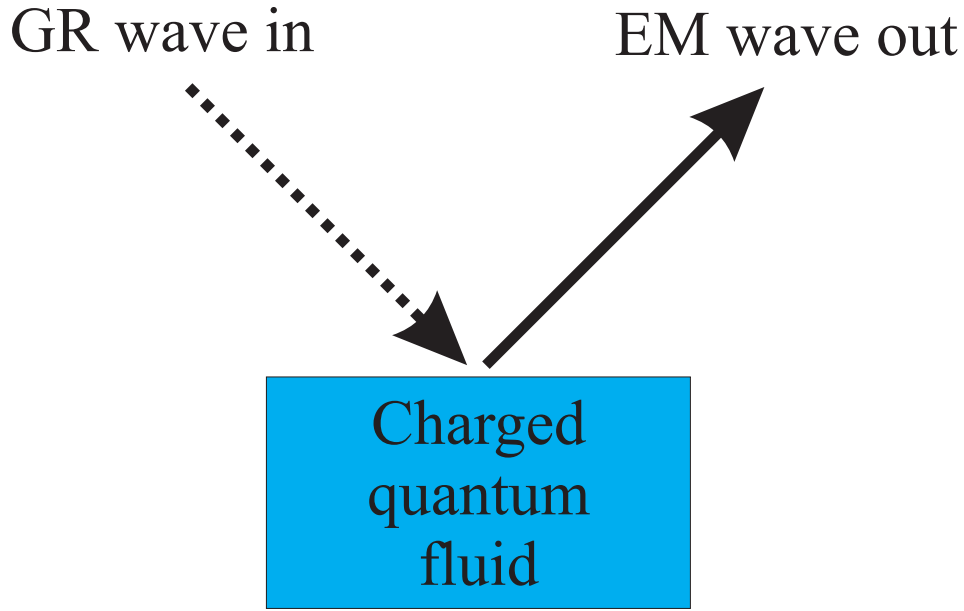}}
\caption{\textquotedblleft Charged quantum fluid\textquotedblright\ is a
quantum transducer consisting of a pair of \textquotedblleft Millikan oil
drops\textquotedblright\ in a strong magnetic field, which converts a
gravity (GR) wave into an electromagnetic (EM) wave.}
\label{GR-to-EM-wave-transducer-4b}
\end{figure}

\begin{figure}[tbp]
\centerline{\includegraphics{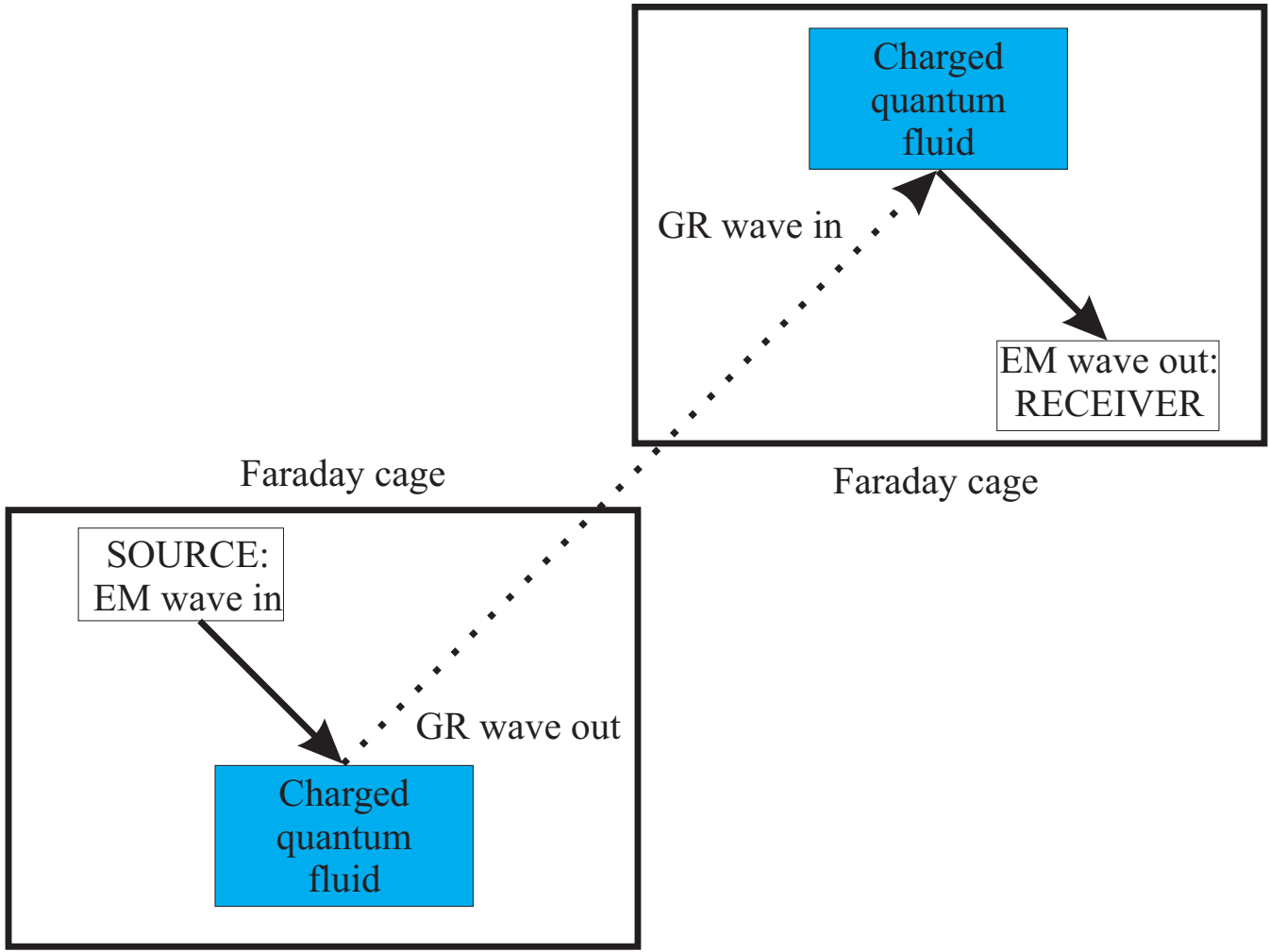}}
\caption{A Hertz-like experiment, in which EM waves are converted by the
lower-left quantum transducer (\textquotedblleft Charged quantum
fluid\textquotedblright ) into GR waves at the source, and the GR waves thus
generated are back-converted back into EM waves by the upper-right quantum
transducer at the receiver. Communication by EM waves is prevented by the
normal (i.e., nonsuperconducting) Faraday cages.}
\label{Hertz-4b}
\end{figure}

This raises the possibility of performing a Hertz-like experiment, in which
the time-reversed quantum transducer process becomes the source, and its
reciprocal quantum transducer process becomes the receiver of GR waves. See
Figure \ref{Hertz-4b}. Faraday cages consisting of nonsuperconducting metals
prevent the transmission of EM waves, so that only GR waves, which can
easily pass through all classical matter such as the normal (i.e.,
dissipative)\ metals of which standard, room-temperature Faraday cages are
composed, are transmitted between the two halves of the apparatus that serve
as the source and the receiver, respectively. Such an experiment would be
practical to perform using standard microwave sources and receivers, since
the scattering cross-sections and the transducer conversion efficiencies of
the two \textquotedblleft Millikan oil drops\textquotedblright\ turn out not
to be too small, as will be shown below.

\section{M{\"{o}}ssbauer-like response of \textquotedblleft Millikan oil
drops\textquotedblright\ in strong magnetic fields to radiation fields}

Let a pair of levitated \textquotedblleft Millikan oil
drops\textquotedblright\ be placed in strong, Tesla-scale magnetic fields,
and let the drops be separated by a distance on the order of a microwave
wavelength, which is chosen so as to satisfy the impedance-matching
condition for a good quadrupolar microwave antenna.

Now let a beam of electromagnetic waves in the Hermite-Gaussian TEM$_{11}$
mode \cite{Yariv1967}, which has a quadrupolar transverse field pattern that
has a substantial overlap with that of a gravitational plane wave, impinge
at a 45$^{\circ }$ angle with respect to the line joining these two charged
objects. Such a mode has been successfully generated using a
\textquotedblleft T\textquotedblright -shape microwave antenna \cite%
{Chiao2004}. As a result of being thus irradiated, the pair of
\textquotedblleft Millikan oil drops\textquotedblright\ will be driven into
relative motion in an anti-phased manner, so that the distance between them
will oscillate sinusoidally with time, according to an observer at infinity.
Thus the simple harmonic motion of the two drops relative to one another (as
seen by this observer) produces a time-varying mass quadrupole moment at the
same frequency as that of the driving electromagnetic wave. This oscillatory
motion will in turn scatter (in a linear scattering process) the incident
electromagnetic wave into gravitational and electromagnetic scattering
channels with comparable powers, provided that the ratio of quadrupolar
radiation powers is that given by Equation (\ref{Larmor-power-ratio}), i.e.,
is of the order of unity, which will be case if the charge-to-mass ratio of
the drops is the same as that of a single electron on a drop with a critical
mass $m_{\text{crit}}$. The reciprocal scattering process will also have a
power ratio of the order of unity. Pairs of large superfluid drops with many
electrons on them can be used as scatterers, as long as their charge-to-mass
ratio is consistent with Equation (\ref{Larmor-power-ratio}).

The M{\"{o}}ssbauer-like response of \textquotedblleft Millikan oil
drops\textquotedblright\ will now be discussed in more detail.\ Imagine what
would happen if one were to replace an electron in the vacuum with a single
electron which is firmly attached to the outside surface of a drop of
superfluid helium in the presence of a strong magnetic field and at ultralow
temperatures, so that the system of the electron and the superfluid,
considered as a single quantum entity, would form a single, macroscopic
quantum ground state \cite{Gigantic-atom}. Such a quantum system can possess
a sizeable gravitational mass. For the case of many electrons attached to a
large, massive drop, where a quantum Hall fluid forms on the outside surface
of the drop in the presence of a strong magnetic field, there results a
Laughlin-like ground state, which is the many-body state of an
incompressible quantum fluid \cite{Laughlin}. The property of quantum
incompressibility of such a fluid is equivalent to the property of
\textquotedblleft quantum rigidity,\textquotedblright\ which is the first
necessary requirement for achieving high efficiency in
gravitational-radiation antennas, as was pointed out in \cite{Chiao2004}.
Like superfluids and superconductors, this fluid is also dissipationless.
This fulfills the conditon of \textquotedblleft quantum
dissipationlessness,\textquotedblright\ which is the second necessary
requirement for the successful construction of efficient gravity-wave
antennas \cite{Chiao2004}.

In the presence of strong, Tesla-scale magnetic fields, an electron is
prevented from moving at right angles to the local magnetic field line
around which it is executing tight cyclotron orbits. The result is that the
surface of the drop, to which the electron is tightly bound, cannot undergo
low-frequency liquid-drop deformations, such as the oscillations between the
prolate and oblate spheroidal configurations of the drop which would occur
at low frequencies in the absence of the magnetic field. After the drop has
been placed into Tesla-scale magnetic fields at milli-Kelvin-scale operating
temperatures, both the single- and many-electron drop systems will be
effectively frozen into the ground state, since the characteristic energy
scale for electron cyclotron motion in Tesla-scale fields is on the order of
Kelvins. Due to the tight coupling of the electron(s) to the outside surface
of the drop, also on the scale of Kelvins, this would effectively freeze out
all low-frequency shape deformations of the superfluid drop.

Since all internal degrees of freedom of the drop, such as its microwave
phonon excitations, will also be frozen out at sufficiently low
temperatures, the charge and the entire mass of the \textquotedblleft
Millikan oil drop\textquotedblright\ will co-move rigidly together as a
single unit, in a zero-phonon, M\"{o}ssbauer-like response to applied
radiation fields with frequencies below the cyclotron frequency. This is a
result of the elimination of all internal degrees of freedom by the
Boltzmann factor at sufficiently low temperatures, so that the system stays
in its ground state, and only the external degrees of freedom of the drop,
consisting only of its center-of-mass motions, remain.

The criterion for this zero-phonon, or M\"{o}ssbauer-like, mode of response
of the electron-drop system is that the temperature of the system is
sufficiently low, so that the probability for the entire system to remain in
its ground state without even a single quantum of excitation of any of its
internal degrees of freedom being excited, is very high, i.e.,%
\begin{equation}
\text{Prob. of zero internal excitation}\approx 1-\exp \left( -\frac{E_{%
\text{gap}}}{k_{B}T}\right) \rightarrow 1\text{ as }\frac{k_{B}T}{E_{\text{%
gap}}}\rightarrow 0,  \label{Prob(no excitation)}
\end{equation}%
where $E_{\text{gap}}$ is the energy gap separating the ground state from
the lowest permissible excited states, $k_{B}$ is Boltzmann's constant, and $%
T$ is the temperature of the system. Then the quantum adiabatic theorem
ensures that the system will stay adiabatically in the ground state of this
quantum many-body system during adiabatic perturbations, such as those due
to weak, externally applied radiation fields with frequencies below the
cyclotron frequency. By momentum conservation, since there are no internal
excitations to take up the radiative momentum transfer, the center of mass
of the entire system must undergo recoil in the emission and absorption of
radiation. Thus the mass involved in the response to radiation fields is the
entire mass of the whole system.

For the case of a single electron (or many electrons in the case of the
quantum Hall fluid)\ in a strong magnetic field, the typical energy gap is
given by%
\begin{equation}
E_{\text{gap}}=\hbar \omega _{\text{cycl}}=\frac{\hbar eB}{mc}>>k_{B}T\text{
,}  \label{Cyclotron-gap}
\end{equation}%
where $\omega _{\text{cycl}}=eB/mc$ is the electron cyclotron frequency.
This inequality is valid for the Tesla-scale fields and milli-Kelvin-scale
temperatures in the experiments being considered here.

\section{Estimate of the scattering cross-section}

Let $d\sigma _{a\rightarrow \beta }$ be the differential cross-section for
the scattering of a mode $a$ of radiation of an incident gravitational wave
to a mode $\beta $ of a scattered electromagnetic wave by a pair of
\textquotedblleft Millikan oil drops\textquotedblright\ (Latin subscripts
denote GR waves, and Greek subscripts EM waves). Then, by time-reversal
symmetry%
\begin{equation}
d\sigma _{a\rightarrow \beta }=d\sigma _{\beta \rightarrow a}\text{ .}
\end{equation}%
Since electromagnetic and weak gravitational fields both formally obey
Maxwell's equations (apart from a difference in the signs of the source
density and the source current density; see Equations (\ref%
{Maxwell-like-eq-1}) - (\ref{Maxwell-like-eq-4})), and since these fields
obey the same boundary conditions, the solutions for the modes for the two
kinds of scattered radiation fields must also have the same mathematical
form. Let $a$ and $\alpha $ be a pair of corresponding solutions, and $b$
and $\beta $ be a different pair of corresponding solutions to Maxwell's
equations for GR and EM modes, respectively. For example, $a$ and $\alpha $
could represent incoming plane waves which copropagate in the same
direction, and $b$ and $\beta $ scattered, outgoing plane waves which
copropagate together in a different direction. Then for a pair of drops with
the same charge-to-mass ratio as that for critical-mass drops with single
electrons, there is an equal conversion into the two types of scattered
radiation fields in accordance with Equation (\ref{Larmor-power-ratio}), and
therefore%
\begin{equation}
d\sigma _{a\rightarrow b}=d\sigma _{a\rightarrow \beta }\text{ ,}
\end{equation}%
where $b$ and $\beta $ are corresponding modes of the two kinds of scattered
radiations.

By the same line of reasoning, for this pair of drops%
\begin{equation}
d\sigma _{b\rightarrow a}=d\sigma _{\beta \rightarrow a}=d\sigma _{\beta
\rightarrow \alpha }\text{ .}
\end{equation}%
It therefore follows from the principle of reciprocity (i.e., detailed
balance or time-reversal symmetry) that%
\begin{equation}
d\sigma _{a\rightarrow b}=d\sigma _{\alpha \rightarrow \beta }.
\end{equation}

In order to estimate the size of the total cross-section, it is easier to
consider first the case of electromagnetic scattering, such as the
scattering of microwaves from a pair of large drops with radii $R$ and a
separation $r$ on the order of a microwave wavelength (but with $r>2R$). The
diameter $2R$ of the drops can be made to be comparable to their separation $%
r\simeq \lambda $, (e.g., with $2\pi R=\lambda $ for the first Mie
resonance), provided that many electrons are added on their surfaces, so
that their charge-to-mass ratio is maintained to be the same as that of a
single electron on a critical-mass drop (this requires the addition of 20
thousand electrons for the first Mie resonance at $\lambda =2.5$ cm, where $%
R=4$ mm), and therefore Equation (\ref{Larmor-power-ratio}) still holds for
these large drops.

For an incident EM wave of a particular circular polarization, even just a
single, delocalized electron in the presence of a strong magnetic field is
enough to produce specular reflection of this wave (see Appendix A).
Therefore for circularly polarized light, the two drops behave like
perfectly conducting, shiny, mirrorlike spheres, which scatter light in a
manner similar to that of perfectly elastic hard-sphere scattering in
idealized billiards. The total cross section for the scattering of
electromagnetic radiation from this pair of large drops is therefore given
approximately by the geometric cross-sectional areas of two hard spheres%
\begin{equation}
\sigma _{\alpha \rightarrow \text{all }\beta }=\int d\sigma _{\alpha
\rightarrow \beta }\simeq \text{Order of }\pi R^{2}
\label{Geometric-X-section}
\end{equation}%
where $R$ is the hard-sphere radius of a drop. This hard-sphere
cross-section is much larger than the Thomson cross-section for the
classical, \emph{localized} single free-electron scattering of
electromagnetic radiation.

However, if, as one might expect on the basis of the prevailing (but
possibly incorrect) opinion that all gravitational interactions with matter,
including the scattering of gravitational waves from all types of matter, is
completely independent of whether this matter is classical or
quantum-mechanical in nature on any scale of size, and that therefore the
scattering cross-section for the drops would be extremely small as it is for
the classical Weber bar, then by reciprocity, the total cross-section for
the scattering of electromagnetic waves from the two-drop system must also
be extremely small. In other words, if \textquotedblleft Millikan oil
drops\textquotedblright\ were to be essentially invisible to gravitational
radiation as is commonly believed, then by reciprocity they must also be
essentially invisible to electromagnetic radiation. To the contrary, if it
should turn out that the quantum Hall fluid on the surface of these drops
should make them behave like superconducting spheres, then the earlier
discussion in connection with Equation (\ref%
{Reflection-from-vacuum-superconductor-interface}) would imply that the
total cross-section of these drops will be like that of hard-sphere
scattering, so that they certainly would not be invisible.

\begin{figure}[tbp]
\centerline{\includegraphics{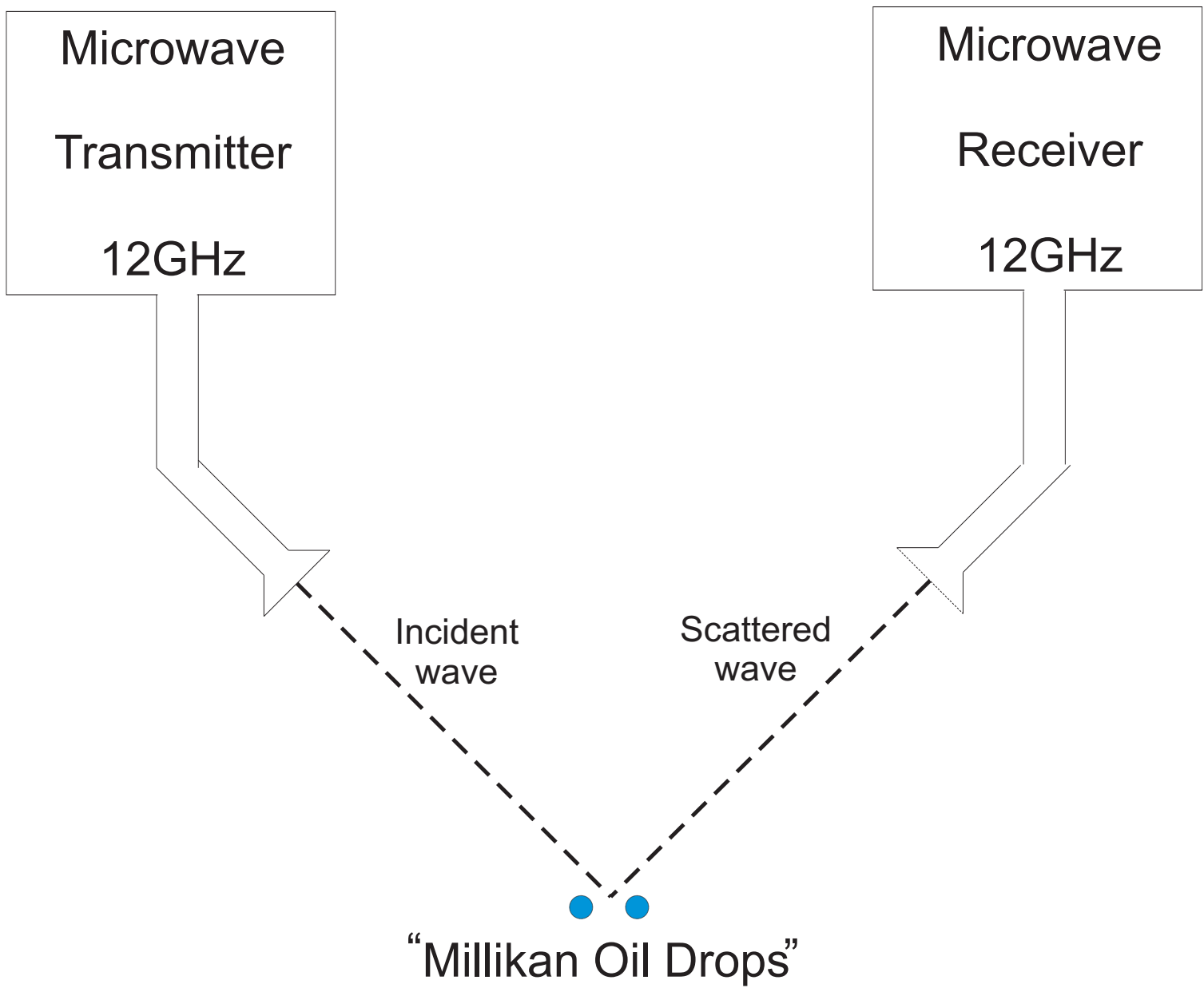}}
\caption{Schematic of apparatus to measure the scattering cross-section of
quadrupolar microwaves from a pair of \textquotedblleft Millikan oil
drops\textquotedblright\ (not to scale).}
\label{Rescaled-scattering-experiment-4}
\end{figure}

In order to check the above hard-sphere scattering cross-section result, we
propose to first perform in a preliminary experiment a measurement of the
scattering cross section for quadrupolar microwave radiation off of a pair
of large \textquotedblleft Millikan oil drops\textquotedblright\ (see Figure %
\ref{Rescaled-scattering-experiment-4}). A standard oscillator at 12 GHz
emits microwaves which are prepared in a quadrupolar TEM$_{11}$ mode and
directed in a beam towards these drops. \ The intensity of the scattered
microwave beam generated by the pair of drops is then measured by means of a
standard 12 GHz heterodyne receiver, which receives a quadrupolar TEM$_{11}$
mode. The purpose of this experiment is the check if the scattering
cross-section is indeed as large as the geometric cross-section predicted by
Equations (\ref{Reflection-from-vacuum-superconductor-interface}) and (\ref%
{Geometric-X-section}). As one increases the temperature, one should observe
the disappearance of this enhanced scattering cross section above the
quantum Hall transition temperature or the superfluid lambda point,
whichever comes first.

\section{A common misconception corrected}

In connection with the idea that an EM wave incident on a pair of drops
could generate a GR wave, there arises a common misconception that the drops
are so heavy that their large inertia will prevent them from moving with any
appreciable amplitude in response to the driving EM wave amplitude. How can
they then possibly generate copious amounts of GR waves? This objection
overlooks the major role played by the principle of equivalence in the
motion of the drops, as will be explained below.

According to the equivalence principle, two tiny inertial observers, who are
undergoing free fall, i.e., who are freely floating near their respective
centers of the two \textquotedblleft Millikan oil drops,\textquotedblright\
would see no acceleration at all of the nearby surrounding matter of their
drop (nor would they feel any forces) due to the gravitational fields
arising from a gravity wave passing over the two drops. However, when they
measure the distance separating the two drops, by means of laser
interferometry, for example, they would conclude that the other drop is
undergoing acceleration relative to their drop, due to the fact that the 
\textit{space} between the drops is being alternately stretched and squeezed
by the incident gravity wave.\ They would therefore further conclude that
the charges attached to the surfaces of their locally freely-falling drops
would radiate electromagnetic radiation, in agreement with the observations
of the observer at infinity, who sees two charges undergoing time-varying
relative acceleration in response to the passage of the gravity wave.

According to the reciprocity principle, this scattering process can be
reversed in time. Under time reversal, the scattered electromagnetic wave
now becomes a wave which is incident on the drops. Again, the two tiny
inertial observers near the center of the drops would see no acceleration at
all of the surrounding matter (nor would they feel any forces) due to the
electric and magnetic fields of the incident electromagnetic wave. Rather,
they would conclude from measurements of the distance separating the two
drops, that it is again the \textit{space} between the drops that is being
alternately squeezed and stretched by the incident electromagnetic wave.
They would again further conclude that the masses associated with their
locally freely-falling drops would radiate gravitational radiation, in
agreement with the observations of the observer at infinity, who sees two
masses undergoing time-varying relative acceleration in response to the
passage of the electromagnetic wave.

From this general relativistic viewpoint, which is based upon the
equivalence principle, the fact that the drops might possess very large
inertias is irrelevant, since in fact the drops are not moving at all with
respect to the local inertial observer located at the center of drop.
Instead of causing motion of the drops \emph{through} space, the
gravitational fields of the incident gravitational wave are acting directly 
\emph{upon} space itself by alternately stretching and squeezing the space
in between the drops. Likewise, in the reciprocal process the very large
inertias of the drops are again irrelevant, since the electromagnetic wave
is not producing any motion at all of these drops with respect to the same
inertial observer \cite{Footnote-B}. Instead of causing motion of the drops 
\emph{through} space, the electric and magnetic fields of the incident
electromagnetic wave are again acting directly \emph{upon} space itself by
alternately squeezing and stretching the space in between the drops. The
time-varying, accelerated motion of the drops as seen by the distant
observer that causes quadrupolar radiation to be emitted in both cases, is
due to the time-varying \textit{curvature} of spacetime induced both by the
incident gravitational wave and by the incident electromagnetic wave. It
should be remembered that the space inside which the drops reside is
therefore no longer flat, so that the Newtonian concept of a
radiation-driven, local accelerated motion of a drop through a fixed, flat
Euclidean space, is therefore no longer valid.

\section{The strain of space produced by the drops for a milliwatt of GR
wave power}

Another common objection to these ideas is that the strain of space produced
by a milliwatt of an electromagnetic wave is much too small to detect.
However, in the Hertz-like experiment, one is not trying to detect directly
the \textit{strain} of space, but rather the \textit{power} that is being
transferred by the gravitational radiation fields from the source to the
receiver.

Let us put in some numbers. Suppose that one succeeded in completely
converting a milliwatt of EM wave power into a milliwatt of GR wave power at
the source. How big a strain amplitude of space would be produced by the
resulting GR wave? The gravitational analog of the time-averaged Poynting
vector is given by \cite{Weinberg}%
\begin{equation}
\left\langle S\right\rangle =c\left\langle t_{\mu \nu }\right\rangle =\frac{%
\omega ^{2}c^{3}}{8\pi G}h_{+}^{2}
\end{equation}%
where $\left\langle t_{\mu \nu }\right\rangle $ are certain components of
the time-averaged stress-energy tensor of a plane wave and $h_{+}$ is the
dimensionless strain amplitude of space for one polarization of a
monochromatic plane wave. For a milliwatt of power in such a plane wave at
30 GHz focused by means of a Newtonian telescope to a 1 cm$^{2}$ Gaussian
beam waist, one obtains a dimensionless strain amplitude of%
\begin{equation}
h_{+}\simeq 2\times 10^{-24}.
\end{equation}%
This strain is indeed exceedingly difficult to directly detect. \ However,
it is not necessary to directly measure the strain of space in order to
detect gravitational radiation, just as it is not necessary to directly
measure the \emph{electric field} of a light wave, which may also be
exceedingly small, in order to be able to detect this wave. \ Instead, one
can measure directly the \emph{power} conveyed by a beam of light by means
of bolometry, for example. \ Likewise, if one were to succeed to completely
back-convert this milliwatt of GR wave power back into a milliwatt of EM
power at the receiver, this amount of power would be easily detectable by
standard microwave techniques.

\section{Signal-to-noise considerations}

The signal-to-noise ratio expected for the Hertz-like experiment depends on
the current status of microwave source and receiver technologies. Based on
the experience gained from the experiment done on YBCO using existing
off-the-shelf microwave components \cite{Chiao2004}, we expect that we would
need geometric-sized cross-sections and a minimum conversion efficiency on
the order of a few parts per million per transducer, in order to detect a
signal. The overall signal-to-noise ratio depends on the initial microwave
power, the scattering cross-section, the conversion efficiency of the
quantum transducers, and the noise temperature of the microwave receiver
(i.e., its first-stage\ amplifier).

Microwave low-noise amplifiers can possess noise temperatures that are
comparable to room temperature (or even better, such as in the case of
liquid-helium cooled paramps or masers used in radio astronomy). The minimum
power $P_{\min }$ detectable in an integration time $\tau $ is given by%
\begin{equation}
P_{\min }=\frac{k_{B}T_{\text{noise}}\Delta \nu }{\sqrt{\tau \Delta \nu }}
\end{equation}%
where $k_{B}$ is Boltzmann's constant, $T_{\text{noise}}$ is the noise
temperature of the first stage microwave amplifier, and $\Delta \nu $ is its
bandwidth. Assuming an integration time of one second, and a bandwidth of 1
GHz, and a noise temperature $T_{\text{noise}}=300$ K, one gets $P_{\min
}(\tau =$1 sec$)=1.3\times 10^{-25}$ Watts, which is much less than the
milliwatt power levels of typical microwave sources.

\section{Possible applications}

\bigskip If we should be successful in the Hertz-like experiment, this could
lead to important possible applications in science and engineering. In
science, it would open up the possibility of gravity-wave astronomy at
microwave frequencies. One important problem to explore would be
observations of the analog of the Cosmic Microwave Background (CMB) in
gravitational radiation. Since the Universe is much more transparent to
gravity waves than to electromagnetic waves, such observations would allow a
much more penetrating look into the extremely early Big Bang towards the
Planck scale of time, than the presently well-studied CMB. \ Different
cosmological models of the very early Universe give widely differing
predictions of the spectrum of this penetrating radiation, so that by
measurements of the spectrum, one could tell which model, if any, is close
to the truth \cite{NASA2006}. The anisotropy in this radiation would also be
very important to observe.

In engineering, it would open up the possibility of intercontinental
communication by means of microwave-frequency gravity waves directly through
the interior of the Earth, which is transparent to such waves. This would
eliminate the need of communications satellites, and would allow
communication with people deep underground or underwater in submarines in
the Oceans. Such a new direction of gravity-wave engineering could aptly be
called \textquotedblleft gravity radio\textquotedblright .

\section{Appendix A: specular reflection of a circularly polarized EM wave
by a delocalized electron moving on a plane in the presence of a strong
magnetic field}

Here we address the question: What is the critical frequency for specular
reflection of an EM plane wave normally incident upon a plane, in which
electrons are moving in the presence of a strong B field?\ \ The motivation
for solving this problem is to answer also the following questions: How can
just a single electron on the outside surface of a \textquotedblleft
Millikan oil drop\textquotedblright\ generate enough current in response to
an incident EM wave, so as to produce a re-radiated wave which totally
cancels out the incident wave within the interior of the drop, with the
result that none of the incident radiation can enter into the drop? Why does
specular reflection occur from the surface of such a drop, and hence why
does a hard-sphere EM cross-section result for a pair of \textquotedblleft
Millikan oil drops\textquotedblright ?

To simplify this problem to its bare essentials, let us examine first a
simpler, planar problem consisting of a uniform electron gas moving
classically on a planar dielectric surface. We shall start from a 3D point
of view, but the Coulombic attraction of the electrons to their image
charges inside the dielectric will confine them in the direction normal to
the plane, so that the electrons are restricted to a 2D motion, i.e., to
motion in the two transverse dimensions of the plane. The electrons are
subjected to a strong DC magnetic field applied normally to this plane. What
is the linear response of this electron gas to a weak, normally incident EM
plane wave? Does a specular plasma-like reflection occur below a critical
frequency, even when just only a single, delocalized electron is present on
the plane? Let us first solve this problem classically.

Let the plane in question be the $z=0$ plane, and let a strong, applied DC $%
\mathbf{B}$ field be directed along the positive $z$ axis. The Lorentz force
on an electron is given by%
\begin{equation}
\mathbf{F}=e\left( \mathbf{E}+\frac{\mathbf{v}}{c}\mathbf{\times B}\right)
\end{equation}%
where $\mathbf{E}$, the weak electric field of the normally incident plane
wave, lies in the $(x,y)$ plane. (We shall use Gaussian units here.) The
cross product $\mathbf{v\times B}$ is given by%
\begin{equation}
\mathbf{v\times B=}\left\vert 
\begin{array}{ccc}
\mathbf{i} & \mathbf{j} & \mathbf{k} \\ 
v_{x} & v_{y} & 0 \\ 
0 & 0 & B%
\end{array}%
\right\vert =\mathbf{i}v_{y}B-\mathbf{j}v_{x}B~.
\end{equation}%
Hence Newton's equations of motion reduce to $x$ and $y$ components only%
\begin{equation}
F_{x}=m\ddot{x}=eE_{x}+\frac{v_{y}}{c}eB=eE_{x}+\frac{\dot{y}}{c}eB
\label{Newton-x}
\end{equation}%
\begin{equation}
F_{y}=m\ddot{y}=eE_{y}-\frac{v_{x}}{c}eB=eE_{x}-\frac{\dot{x}}{c}eB\text{ .}
\label{Newton-y}
\end{equation}%
Let us assume that the driving plane wave is a weak monochromatic wave with
the exponential time dependence%
\begin{equation}
E=E_{0}\exp \left( -i\omega t\right) ~.
\end{equation}%
Then assuming a linear response of the system to the weak incident EM wave,
the displacement, velocity, and acceleration of the electron all have the
same exponential time dependence%
\begin{equation}
x=x_{0}\exp \left( -i\omega t\right) \text{ and }y=y_{0}\exp \left( -i\omega
t\right)
\end{equation}%
\begin{equation}
\dot{x}=\left( -i\omega \right) x\text{ and }\dot{y}=\left( -i\omega \right)
y
\end{equation}%
\begin{equation}
\ddot{x}=-\omega ^{2}x\text{ and }\ddot{y}=-\omega ^{2}y
\end{equation}%
which converts the two ODEs, Equations (\ref{Newton-x}) and (\ref{Newton-y}%
), into the two algebraic equations for \ $x$ and $y$%
\begin{equation}
-m\omega ^{2}x=eE_{x}-\frac{i\omega y}{c}eB
\end{equation}%
\begin{equation}
-m\omega ^{2}y=eE_{y}+\frac{i\omega x}{c}eB\text{ .}
\end{equation}%
Let us now add $\pm i$ times the second equation to the first equation.
Solving for $x\pm iy$, one gets%
\begin{equation*}
x\pm iy=e\left( \frac{E_{x}\pm iE_{y}}{-m\omega ^{2}\pm \omega eB/c}\right)
\end{equation*}%
where the upper sign corresponds to an incident clockwise circularly
polarized EM, and the lower sign to an anti-clockwise one. Let us define as
a shorthand notation%
\begin{equation}
z_{\pm }\equiv x\pm iy
\end{equation}%
as the complex representation of the displacement of the electron. Solving
for $z_{\pm }$, one obtains%
\begin{equation}
z_{\pm }=\frac{eE_{\pm }}{-m\left( \omega ^{2}\mp \omega \omega _{\text{cycl}%
}\right) }
\end{equation}%
where the cyclotron frequency $\omega _{\text{cycl}}$ is defined as%
\begin{equation}
\omega _{\text{cycl}}\equiv \frac{eB}{mc}\text{ ,}
\end{equation}%
and where%
\begin{equation*}
E_{\pm }\equiv E_{x}\pm iE_{y}\text{ .}
\end{equation*}%
For a gas of electrons with a uniform number density $n_{e}$, the
polarization of this medium induced by the weak incident EM wave is given by%
\begin{equation}
P_{\pm }=n_{e}e\left( x\pm iy\right) =n_{e}ez_{\pm }=\frac{n_{e}e^{2}E_{\pm }%
}{-m\left( \omega ^{2}\mp \omega \omega _{\text{cycl}}\right) }=\chi E_{\pm }
\end{equation}%
where the susceptibility of the electron gas is given by%
\begin{equation}
\chi =\frac{n_{e}e^{2}}{-m\left( \omega ^{2}\mp \omega \omega _{\text{cycl}%
}\right) }=-\frac{\omega _{\text{plas}}^{2}/4\pi }{\omega ^{2}\mp \omega
\omega _{\text{cycl}}}
\end{equation}%
where the plasma frequency $\omega _{\text{plas}}$ is defined by%
\begin{equation}
\omega _{\text{plas}}\equiv \sqrt{\frac{4\pi n_{e}e^{2}}{m}}\text{ .}
\end{equation}%
The index of refraction of the gas $n(\omega )$\ is given by%
\begin{equation}
n(\omega )=\sqrt{1+4\pi \chi (\omega )}=\sqrt{1-\frac{\omega _{\text{plas}%
}^{2}}{\omega ^{2}\mp \omega \omega _{\text{cycl}}}}\text{ .}
\end{equation}%
specular reflection occurs when the index of refraction becomes a pure
imaginary number. Let us define as the critical frequency $\omega _{\text{%
crit}}$ as the frequency at which the index vanishes, which occurs when%
\begin{equation}
\frac{\omega _{\text{plas}}^{2}}{\omega _{\text{crit}}^{2}\mp \omega _{\text{%
crit}}\omega _{\text{cycl}}}=1.
\end{equation}%
Since the index vanishes at this critical frequency, the Fresnel reflection
coefficient ${\mathcal{R}}(\omega )$ from the planar structure for normal
incidence at criticality is given by%
\begin{equation}
{\mathcal{R}}(\omega )=\left\vert \frac{n(\omega )-1}{n(\omega )+1}%
\right\vert ^{2}\rightarrow 100\%\text{ when }\omega \rightarrow \omega _{%
\text{crit}}\text{ ,}
\end{equation}%
which implies specular reflection of the incident plane EM wave from the
electron gas. \ This yields a quadratic equation for $\omega _{\text{crit}}$%
\begin{equation}
\omega _{\text{crit}}^{2}\mp \omega _{\text{crit}}\omega _{\text{cycl}%
}-\omega _{\text{plas}}^{2}=0.
\end{equation}%
The solution for $\omega _{\text{crit}}$ is%
\begin{equation}
\omega _{\text{crit}}=\frac{\pm \omega _{\text{cycl}}\pm \sqrt{\omega _{%
\text{cycl}}^{2}+4\omega _{\text{plas}}^{2}}}{2}.  \label{quadratic-solution}
\end{equation}%
The first $\pm $ sign is physical, and is determined by the sense of
circular polarization of the incident plane wave. The second $\pm $ sign is
mathematical, and originates from the square root. \ One of the latter
mathematical signs is unphysical. To determine which choice of the latter
sign is physical and which is unphysical, let us first consider the limiting
case when the inequality%
\begin{equation}
\omega _{\text{cycl}}<<\omega _{\text{plas}}
\end{equation}%
holds. This inequality corresponds physically to the situation when the
magnetic field is very weak, but the electron density is very high, so that
the phenomenon of specular reflection of EM waves with frequencies below the
plasma frequency $\omega _{\text{plas}}$ occurs. Let us therefore take the
limit $\omega _{\text{cycl}}\rightarrow 0$ in the solution (\ref%
{quadratic-solution}). Negative frequencies are unphysical, so that we must
choose the positive sign in front of the surd as the only possible physical
solution. Thus in general it must the case that the physical root of the
quadratic is given by%
\begin{equation}
\omega _{\text{crit}}=\frac{\pm \omega _{\text{cycl}}+\sqrt{\omega _{\text{%
cycl}}^{2}+4\omega _{\text{plas}}^{2}}}{2}.  \label{Physical-root}
\end{equation}%
\qquad

\bigskip Let us now focus on the more interesting case where the magnetic
field is very strong, but the number density of electrons is very small, so
that the plasma frequency is very low, corresponding to the inequality 
\begin{equation}
\omega _{\text{cycl}}>>\omega _{\text{plas}}\text{ .}
\end{equation}%
There then are two possible solutions, corresponding to clockwise-polarized
and anti-clockwise-polarized EM waves, respectively, viz.%
\begin{equation}
\omega _{\text{crit,1}}=\omega _{\text{cycl}}\text{ and }\omega _{\text{%
crit,2}}=0\text{ .}  \label{2-physical-solutions}
\end{equation}%
Note the important fact that these solutions are independent of the number
density (or plasma frequency) of the electron gas, which implies that even a
very dilute electron gas system can give rise to specular reflection. \ The
fact that these solutions are independent of the number density also implies
that they would apply to the case of an inhomogeneous electron density, such
as that arising for a single delocalized electron confined to the vicinity
of the plane $z=0$ by the Coulombic attraction to its image. Both solutions
of the quadratic equation (\ref{2-physical-solutions}) are now physical
ones, and imply that whether the sense of rotation of the EM polarization
co-rotates or counter-rotates with respect to the magnetic-field--induced
precession of the guiding center motion of the electron around the magnetic
field, determines which sense of circular polarization is transmitted when $%
\omega >$ $\omega _{\text{crit,2}}=0$, or which sense of circular
polarization is totally reflected when $\omega <\omega _{\text{crit,1}%
}=\omega _{\text{cycl}}$, provided that the frequency of the incident
circularly polarized EM wave is less than the cyclotron frequency $\omega _{%
\text{cycl}}$. \ The interesting solution is the one with the non-vanishing
critical frequency, since it implies that there always exists one solution
where there is specular reflection of the EM wave, even when the number
density of electrons is extremely low, i.e., even when the plasma frequency $%
\omega _{\text{plas}}$ approaches zero, and even when this number density
becomes very inhomogeneous as a function of $z$.

In the extreme case of a single electron completely delocalized on the
outside surface of superfluid helium (for example, in an $S$ state on the
outside surface of a spherical drop), one should solve the problem quantum
mechanically, by going back to Landau's solution of the motion of an
electron in a uniform magnetic field, and adding as a time-dependent
perturbation the weak (classical) incident circularly polarized plane wave.
\ However, the above classical solution is sufficient, since it should hold
in the correspondence principle limit, where, for the single delocalized
electron, the effective number density of the above classical solution is
determined by the absolute square of the electron wavefunction, viz.%
\begin{equation}
n_{e}=\left\vert \psi _{e}\right\vert ^{2}\text{ , and}
\end{equation}%
\begin{equation}
\int n_{e}dV=\int \left\vert \psi _{e}\right\vert ^{2}dV=1.
\end{equation}%
Here we must take into account the fact that there is a finite confinement
distance (around 80\AA ) in the $z$ direction of the electron's motion in
the hydrogenic ground state caused by the Coulomb attraction of the electron
to its image charge induced in the dielectric, but the electron is
completely delocalized in the $x$ and $y$ directions on an arbitrarily large
plane (and hence over the large spherical surface of a large drop). The
effective plasma frequency of the single electron may be extremely small;
nevertheless, total reflection by this single, delocalized electron still
occurs, provided that the frequency of the incident circularly polarized EM
wave is below the cyclotron frequency. The fundamental reason why even just
a single electron in a strong magnetic field can give rise to 100\%
reflection is that the $\mathbf{v\times B}$ Lorentz force leads to an
arbitrarily large transverse induced current (i.e., to an arbitrarily large
Hall \emph{conductivity} of the electron on the plane, even though the
quantum Hall \emph{conductance} $e^{2}/h$ may be finite), which effectively
shorts out the incident circularly polarized EM wave. Thus one concludes
that the hard-wall boundary conditions used in the order-of-magnitude
estimate given by Equation (\ref{Geometric-X-section}) of the scattering
cross-section of microwaves from the drops are reasonable ones. This
conclusion will be tested experimentally (see Figure (\ref%
{Rescaled-scattering-experiment-4})).

\bigskip \textbf{Acknowledgments} I thank John Barrow, Fran\c{c}ois
Blanchette, George Ellis, Sai Ghosh, Dave Kelley, Tom Kibble, Steve Minter,
Kevin Mitchell, James Overduin, Richard Packard, Jay Sharping, Martin
Tajmar, and Roland Winston for their help.


\begin{thebibliography}{99}
\bibitem{MTW} C. W. Misner, K. S. Thorne, and J. A. Wheeler, \textit{%
Gravitation} (Freeman, San Francisco, 1972).

\bibitem{Landau} L.~Landau and E.~Lifshitz, \textit{The Classical Theory of
Fields}, 1st edition (Addison-Wesley, Reading, MA, 1951), page 331, Equation
(11-115).

\bibitem{Weinberg} S. Weinberg, \textit{Gravitation and Cosmology} (John
Wiley \& Sons, New York, 1972).

\bibitem{Taylor1994} J.~G.~Taylor, Rev.~Mod.~Phys.~\textbf{66}, 711 (1994).

\bibitem{Lamb-medal} R. Y. Chiao, J. Mod. Opt. \textbf{53}, 2349 (2006)
(quant-ph/0601193).

\bibitem{Wald} R. M. Wald, \textit{General Relativity} (University of
Chicago Press, Chicago, 1984); V. B. Braginsky, C. M. Caves, and K. S.
Thorne, Phys. Rev. D \textbf{15}, 2047 (1977); A. D. Speliotopoulos and R.
Y. Chiao, Phys. Rev. D \textbf{69}, 084013 (2004). The earliest mention of
Maxwell-like equations for linearized general relativity was perhaps made by
R. L. Forward in Proc. IRE \textbf{49}, 892 (1961).

\bibitem{Kiefer-Weber} C.~Kiefer and C.~Weber, Annalen der Physik (Leipzig) 
\textbf{14}, 253 (2005).

\bibitem{Chiao2004} R.~Y.~Chiao, in \textit{Science and Ultimate Reality},
eds.~J.~D.~Barrow, P.~C.~W.~Davies, and C.~L.~Harper, Jr.~(Cambridge
University Press, Cambridge, 2004), page 254 (quant-ph/0303100); R. Y. Chiao
and W. J. Fitelson (gr-qc/0303089); R. Y. Chiao, W. J. Fitelson, and A. D.
Speliotopoulos (gr-qc/0304026).

\bibitem{Tinkham} M. Tinkham, \textit{Introduction to Superconductivity},
2nd edition (Dover Books on Physics, New York, 2004).

\bibitem{Prange} R. E. Prange and S. M. Girvin, \textit{The Quantum Hall
Effect}, 2nd edition (Springer Verlag, New York, 1990).

\bibitem{Tajmar} M. Tajmar, F. Plesescu, B. Seifert, and K. Marhold,
preprint gr--qc/0610015; M. Tajmar, F. Plesescu, K. Marhold, and C. J. de
Matos, pre-print gr-qc/0603034, (2006); C. J. de Matos, and M. Tajmar,
Physica C \textbf{432}, 167 (2005).

\bibitem{Tajmar2} M. Tajmar, private communication.

\bibitem{Kramers-Kronig} The response of the medium must not only be \emph{%
linear} in the amplitude of the weak applied gravitational radiation fields,
but it must also be \emph{causal}. \ Hence the real and imaginary parts of
the linear response function $\kappa _{G}^{\text{(magn)}}(\omega )$, as a
function of frequency $\omega $ of the gravity wave, must obey
Kramers-Kronig relations similar to those given by Equations (4) and (5) of
Ref. \cite{Chiao2004}(a).

\bibitem{Weilert1996} M.~A.~Weilert, D.~L.~Whitaker, H.~J.~Maris, and
G.~M.~Seidel, Phys.~Rev.~Letters \textbf{77}, 4840 (1996); M.~A.~Weilert,
D.~L.~Whitaker, H.~J.~Maris, and G.~M.~Seidel, J. Low Temp. Phys. \textbf{106%
}, 101 (1997). Another method of levitating charged superfluid drops uses a
parallel-plate capacitor geometry similar to that used by Millikan in his
original oil-drop experiment; see J. J. Niemela, J. Low Temp. Phys. \textbf{%
109}, 709 (1997).

\bibitem{Grimes2} C. C. Grimes and T. R. Brown, Phys. Rev. Letters \textbf{32%
}, 280 (1974); C. C. Grimes and G. Adams, Phys. Rev. Letters \textbf{36},
145 (1976). In the ground state of the system, the electron resides on the 
\emph{outside} surface of a superfluid helium drop, and not within the \emph{%
inside} volume of the drop. When the electron is forced to be within the
interior of the drop, it will form a bubble with a radius of around one
nanometer, due to the balancing of an outwards Pauli pressure with the
surface tension of the superfluid (see R. J. Donnelly, \textit{Experimental
Superfluidity} (The University of Chicago Press, Chicago, 1967), p. 176 ff).
The bubble will then rise to the surface, driven by the Coulomb force of
attraction to its own image charge induced in the surface. It will then
burst through the surface to uniformly coat the drop with one electron
charge on its outside surface. The electron will be in an $S$-state in order
to minimize the energy of the system. This then is the ground state of the
system.

\bibitem{Footnote-A} Note that the quantum-mechanical ground-state
wavefunction (or complex order parameter) must remain \textit{single-valued}
(according to a distant inertial observer) globally at all times everywhere
inside the interior of the system during the passage of a gravity wave. This
is another aspect of the \textquotedblleft quantum
rigidity\textquotedblright\ of a quantum fluid in its response to the
gravity wave.

\bibitem{Yariv1967} A.~Yariv, \textit{Quantum Electronics}, 1st edition
(John Wiley \& Sons, New York, 1967), page 223ff.{}

\bibitem{Gigantic-atom} This single quantum entity can be viewed as if it
were a gigantic atom, in which the usual atomic nucleus is replaced by the
superfluid helium drop, and the usual electronic cloud surrounding the
atomic nucleus is replaced by the electrons on the surface surrounding the
drop. The large energy gap (Equation (\ref{Cyclotron-gap})) arising from the
large applied magnetic field is what makes this gigantic atom extremely
rigid and dissipationless at low temperatures. A pair of such gigantic atoms
forms a gigantic diatomic molecule. If the charges and masses of the two
drops are slightly different from each other, such a gigantic diatomic
molecule will form an entangled state of charge and mass in its ground
state. 

\bibitem{Laughlin} R. B. Laughlin, Phys. Rev. Letters \textbf{50}, 1395
(1983).

\bibitem{Footnote-B} The principle of equivalence should apply to all
charges and fields in curved spacetime \cite{Landau}\cite{Lamb-medal}.
However, Maxwell's equations, as usually formulated for standard
electromagnetism, are expressed in terms of fields on a \emph{flat}
spacetime. They must be generalized to fields on a \emph{curved} spacetime
when interactions with gravitational radiation are considered. The
back-action of EM waves propagating in a curved spacetime upon GR waves can
in principle arise from the contribution of the Maxwell stress-energy
tensor, which is \emph{quadratic} in the EM field strengths, as a source
term on the right-hand side of Einstein's field equations. Such quadratic
terms would give rise to second harmonic generation in the conversion of EM
to GR waves, but not to first harmonic generation. However, there can in
principle arise a \emph{linear} coupling of EM to GR waves when a strong DC
magnetic field is present, and Einstein's equations are linearized in the
weak EM and GR wave amplitudes. This linear coupling arises from a mixing
term, which consists of a product of the DC magnetic field strength and the
EM wave amplitude in the quadratic Maxwell stress-energy tensor that leads
to first harmonic generation of GR waves at the same frequency as that of
the incident EM waves in a linear scattering process. The role of the
\textquotedblleft Millikan oil drops\textquotedblright\ is that they can
greatly enhance the coupling between EM and GR waves due to their extreme
rigidity and large masses. The electrons on their surfaces tightly tie the
local \textbf{B} field lines to these drops, so that these lines are firmly
anchored to the drops. At very low temperatures when the system remains
adiabatically in the ground state, the \textbf{B} field lines and the drops
co-move rigidly together according to a distant observer, when the system is
disturbed by the passage of a GR or an EM wave. A given drop, however,
remains at rest with respect to a local inertial observer at the center of
the drop, and the local \textbf{B} field lines also do not appear to move
with respect to this local inertial observer. By contrast, to the distant
inertial observer in an asymptotically flat region of spacetime far away
from the drops, where radiation fields become asymptotically well defined, a
pair of massive, charged drops appear to be in relative motion, and the
system emits energy in both GR and EM radiations. Thus a graviton (spin 2)
can in principle be produced from a photon (spin 1) in the presence of a DC
magnetic field (spin 1), in a scattering process from the drops.

\bibitem{NASA2006} R. Y. Chiao in the proceedings of the NASA conference
\textquotedblleft Quantum to Cosmos\textquotedblright\ (quant-ph/0606118),
to be published in Int. J. Mod. Phys. D.
\end{thebibliography}
\end{document}